\documentclass[preprint,aps,floatfix,showpacs]{revtex4}

\usepackage{dcolumn}
\usepackage{graphicx}
\usepackage{amsfonts}
\usepackage{amsmath,bm,amssymb}
\usepackage{comment}
\usepackage[usenames]{color}

\begin{document}

\title{Magnetic proximity effect at Bi$_2$Se$_3$/EuS interface with broken inversion symmetry}

\author{Alex Taekyung Lee} \email{techlee@kaist.ac.kr}
\affiliation{Department of Physics, Korea Advanced
Institute of Science and Technology (KAIST), Daejeon 305-701, Korea}

\author{Myung Joon Han} \email{mj.han@kaist.ac.kr}
\affiliation{Department of Physics and KAIST Institute for the
  NanoCentury, Korea Advanced Institute of Science and Technology,
  Daejeon 305-701, Korea }

\author{Kyungwha Park} \email{kyungwha@vt.edu}
\affiliation{Department of Physics, Virginia Tech, Blacksburg, Virginia 24060, USA}

\date{\today}

\begin{abstract}
We investigate a magnetic proximity effect on Dirac surface states of topological insulator
(TI) induced by a Bi$_2$Se$_3$/EuS interface, using density-functional theory (DFT) and
a low-energy effective model, motivated by a recent experimental realization of the interface.
We consider a thin ferromagnetic insulator EuS film stacked on top of Bi$_2$Se$_3$(111) slabs
of three or five quintuple layers (QLs) with the magnetization of EuS normal to the interface
($z$ axis), which breaks both time-reversal and inversion symmetry.
It is found that a charge transfer and surface relaxation makes the Dirac cones electron-doped.
For both 3 and 5 QLs, the top-surface Dirac cone has an energy gap of 9 meV, while the bottom
surface Dirac cone remains gapless. This feature is due to the short-ranged induced magnetic
moment of the EuS film. For the 5 QLs, an additional Dirac cone with an energy gap
of 2 meV is formed right below the bottom-surface Dirac point, while for 3 QLs, there is no
additional Dirac cone. We also examine the spin-orbital texture of the Dirac surface states
with broken inversion symmetry, using DFT and the effective model.
We find that the $p_z$ orbital
is coupled to the $z$ component of the spin moment in the opposite sign to the $p_x$ and $p_y$
orbitals. The $p_z$ and radial $p$ orbitals are coupled to the in-plane spin texture
in the opposite handedness to the tangential $p$ orbital. The result obtained from the effective
model agrees with our DFT calculations. The calculated spin-orbital texture may be tested from
spin-polarized angle-resolved photoemission spectroscopy.

\end{abstract}

\pacs{73.20.-r, 73.20.At, 71.15.Mb}

\maketitle

\section{Introduction}

Topological insulators (TIs) are distinguishable from ordinary insulators in that Dirac
surface states are topologically protected by time-reversal symmetry and that they have helical
spin texture \cite{Hasan,Qi_Zhang-RMP}. When TIs are interfaced with magnetic substrates or doped
with magnetic elements, time-reversal symmetry can be broken. As a result, not only an energy gap
is open at the Dirac point \cite{YLChen}, but also various interesting effects are
expected, such as the topological magneto-electric effect \cite{Qi_Zhang-RMP}, the quantum anomalous
Hall effect \cite{Yu-QAH,Chang-QAL}, weak localization behavior in transport
\cite{Liu-WAL,He-WAL,Shen-WAL1}, and enhanced spin transfer
torque \cite{Mahfouzi,EAKim}.


Despite the previous studies on magnetic TIs, there remains several important issues that need to be resolved.
First, when a TI is in contact with a ferromagnetic
material, the coupling between them may not be weak. Consequently, the magnetic material may become
nonmagnetic at the interface \cite{Li-Co} or the magnetic substrate bands can be dominant near the Fermi
level with the absence of TI surface states \cite{Luo-MnSe}. Second, when magnetic atoms such as Fe or
Cr were doped in a TI \cite{Wray,Valla,Li-Co,Chang-Cr}, it was reported that the dopants may form clusters
rather than being uniformly distributed without ferromagnetic ordering \cite{Chang-Cr}. Last, the magnetic
easy axis of a magnetic TI may not be perpendicular to the surface \cite{Yang-easy_axis}.

Considering these issues, an interface between a ferromagnetic insulator (FMI) and a TI with a magnetic
easy axis normal to the surface, is a good candidate system to study the magnetic proximity effect
\cite{Luo-MnSe,Chulkov-1,Wei-EuS,Yang-EuS}.
In this regard, a superlattice of Bi$_2$Se$_3$ (TI) interfaced with MnSe (FMI) in the presence of
inversion symmetry (IS) was theoretically studied \cite{Luo-MnSe,Chulkov-1}.
In this case, the Dirac surface states identified in Ref.\cite{Luo-MnSe} are not truly TI surface states
because the TI surface-surface hybridization greatly increases with decreasing the FMI thickness. The
large contamination of the Dirac surface states by the FMI is consistent with the result in Ref.\cite{Chulkov-1}.
In addition, the Bi$_2$Se$_3$/MnSe interface has not been experimentally realized yet, and it has a lattice
mismatch of 1.9 \%. Recently, the Bi$_2$Se$_3$/EuS interface has been fabricated and their transport properties have
been measured \cite{Wei-EuS,Yang-EuS}, where the lattice mismatch is less than 1~\% and the structure has
broken IS with the magnetic easy axis normal to the interface ($z$ axis). There are no theoretical studies
of the magnetic proximity effect at this interface yet.

One interesting feature of the TI arising from strong spin-orbit coupling (SOC) is that the orbitals forming the topological surface
states are highly coupled to a particular spin texture. This spin-orbital texture were studied for pristine TIs
\cite{Zhang-spin,Zhu-spin,CAO13}, and it was shown to be controlled by the polarization of incident photons
\cite{PARK_LOUIE_12,Jozwiak}. Therefore, an interesting question is how the spin-orbital texture
of the surface states is modified by the magnetic proximity effect.

In this work, we present the spin-orbital texture and electronic structure of the surface states
induced by the magnetic proximity effect at the Bi$_2$Se$_3$/EuS interface, using first-principles and effective
model calculations. We consider a slab geometry where an EuS film with magnetic moment along the $z$ axis is
placed on top of Bi$_2$Se$_3$, which breaks both IS and time-reversal symmetry. The first-principles calculations
show that three massive Dirac cones are present, and that EuS-dominated bands do not appear near
the Fermi level. Among the three Dirac cones, the states localized into the top (bottom) TI surface have an energy
gap of 9~meV (less than 1 meV), while a new Dirac cone slightly deeper into the top TI surface has an energy gap of
2 meV. By constructing a low-energy effective model for the surface states, we find that at small momentum the $p_z$
orbital is coupled to the $z$ component of the spin moment in the opposite sign to the $p_x$ and $p_y$ orbitals,
while the $p_z$ and radial $p$ orbitals are coupled to the in-plane spin texture in the opposite handedness to
the tangential $p$ orbital. This result agrees with our
first-principles calculations of the spin-orbital texture. The calculated spin-orbital texture can be observed
from spin-polarized angle-resolved photoemission spectroscopy (ARPES).

\section{First-principles calculation method}

We perform density-functional theory (DFT) calculations, using the projector augmented wave (PAW) potentials
\cite{PAW} and generalized gradient approximation (GGA) \cite{PBE} as an exchange-correlation functional,
implemented in the \textsf{VASP} code \cite{VASP}. We consider SOC self-consistently,
where the spin quantization axis is the $z$ axis. All Eu $f$ electrons are taken as valence
electrons. To include an additional correlation effect of Eu, we use $U_f$=8 eV and $J_f$=1 eV
for Eu $f$ orbitals within the GGA+$U$ scheme \cite{LDA+U}. Then we find that the optimized lattice
constant of bulk EuS is 6.014~\AA,~and that the band gap is 1.12~eV comparable to the experimental
value of 1.65 eV \cite{EuX,Ghosh}.

\begin{figure}
\begin{center}
\includegraphics[width=12cm, angle=0]{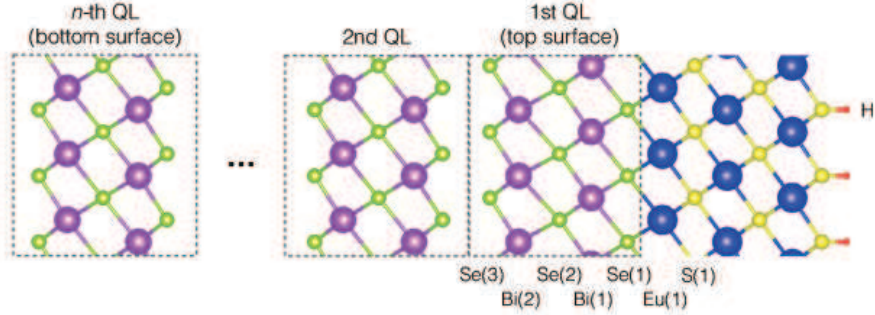}
\caption{ (Color online) Schematic geometry of the Bi$_2$Se$_3$(111)/EuS(111) interface.
The states localized into the 1st QL or the topmost QL are referred to as top-surface states.
}
\label{fig:geo}
\end{center}
\end{figure}

We consider a supercell where an EuS(111) film is on a Bi$_2$Se$_3$(111) slab of 3 or 5 quintuple
layers (QLs) with a vacuum layer thicker than 40~\AA. Note that in this slab geometry both IS and time-reversal
symmetry are broken. With broken IS, the top-surface and bottom-surface Dirac cones are separated, resulting in
gapless Dirac dispersion even for a Bi$_2$Se$_3$ slab as thin as 2 QLs \cite{Park13}. This TI/FMI slab geometry
can be taken as a "model" for experimental systems, considering that non-collinear magnetization and magnetic
domains in EuS \cite{Wei-EuS,Yang-EuS} are not included. However, in the first-principles calculations, the
coupling between the topological surface states and bulk-like states is inherently included as well as the
coupling between surfaces, in contrast to the effective model for the surfaces \cite{Yu-QAH}.
At the Bi$_2$Se$_3$/EuS interface, we confirm that Se-Eu bonding is favorable over Se-S bonding, as expected from
an ionic character of EuS and due to nominal valence states of Bi and Se. The topmost S atom in the Bi$_2$Se$_3$/EuS
slab is passivated with H to avoid dangling bonds.
We consider two configurations of the EuS film, such as Eu(1) at a fcc or hcp site of the TI slab.
Relaxation of the slab geometry is carried out for both EuS configurations, until the
forces on the EuS film and on the top four atomic layers in the TI slab (Bi(1), Se(1), Bi(2), Se(2) in
Fig.~\ref{fig:geo}), are less than 0.01~eV/\AA,~while the rest of the atoms are fixed with the experimental values
obtained from the experimental lattice constant of Bi$_2$Se$_3$, $a$=4.143\AA~\cite{NAKA63}.
We find that  the fcc site gives an energy lower than the hcp site by 47 meV.
 Thus, henceforth, we consider only the fcc site for our
calculations of band structures. The optimum Eu-Se bond length turns out to be 3.059~\AA, which is
similar to that in the bulk rock salt EuSe \cite{EuX}. Self-consistent calculations
on the relaxed geometries are performed with an energy cutoff of 500~eV and 9$\times$9$\times$1 $k$-points
until the total energy converges to $10^{-5}$~eV.
Regarding dipole corrections, for 3 QLs, we confirm that they do not
affect the band structure as long as the vacuum is thick enough.

\section{Electronic structure of Bi$_2$Se$_3$/EuS}

\begin{figure}
\begin{center}
\includegraphics[width=10cm, angle=0]{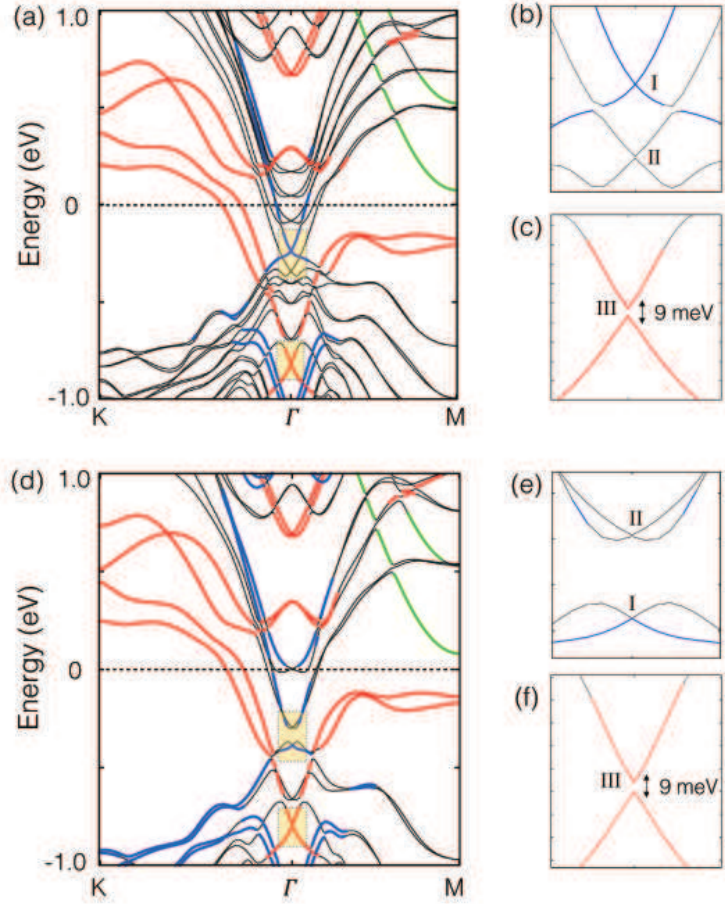}
\caption{(Color online) DFT-calculated band structure of the Bi$_2$Se$_3$/EuS slab when the TI slab
is (a) 5 QLs or (d) 3 QLs, where the top-surface and bottom-surface states are marked in red and blue,
respectively. Here the states in green are EuS bands, and the states in black are bulk-like states.
The definitions of the different types of the states can be found in the main text. The Fermi level
$E_F$ shown as the dashed lines are set to zero in energy. Several bands in the vicinity of the
bottom-surface and top-surface Dirac points for the 5 QLs [(b),(c)] and
the 3 QLs [(e),(f)], are highlighted. }
\label{fig:band}
\end{center}
\end{figure}

\begin{figure}
\begin{center}
\includegraphics[width=10cm, angle=0]{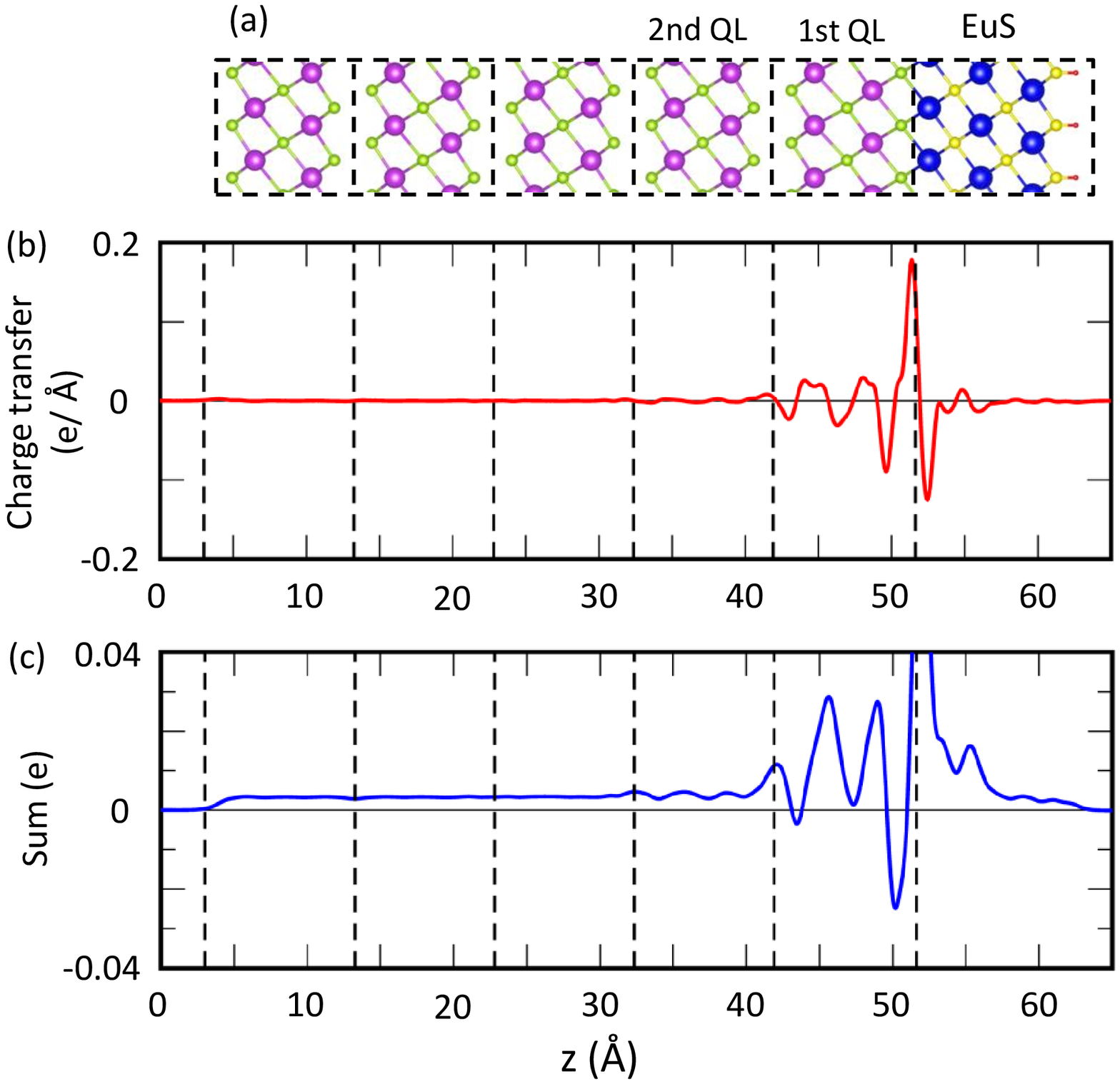}
\caption{(Color online) (a) 5QLs/EuS slab geometry (b) Charge transfer $C_{\textmd{net}}$ as a function
of $z$ for the 5QL/EuS slab. (c) Cumulative sum of the charge transfer vs $z$ for the 5QL/EuS slab.
The vertical dashed lines separate adjacent QLs.}
\label{fig:chg}
\end{center}
\end{figure}

\subsection{Results of 5-QL Bi$_2$Se$_3$}

Figure~\ref{fig:band}(a) shows the calculated band structure of the 5QL/EuS slab where the top and bottom
TI surface states are marked in red and blue, and EuS bands in green. The top and bottom-surface states are
the ones where more than 40\% of the electron density is localized into the topmost and bottommost QLs,
respectively \cite{Park10}. EuS-dominated states are defined as states with more than 30\% of the density
onto the EuS film. The exact percentages would not change the identification of the projected bands.
A TI slab of 5 QLs has a weak surface-surface hybridization, considering that the decay
length of the TI surface states is about 2 QLs ($\sim$2 nm).

The EuS bands are located outside the TI bulk band gap, and they are well separated from the top-surface
and bottom-surface Dirac cones, as shown in Fig.~\ref{fig:band}(a).
This is a distinctive feature of a TI/FMI
interface and differs from the cases of metallic substrates or high concentrations of dopants or adatoms. One
caveat is that two EuS bands appear close to the Fermi level $E_F$ near the $\overline{M}$ point.
These bands are expected to move further upward with on-site Coulomb repulsion for Eu $d$ electrons
($U_d$) \cite{Ghosh}.

The calculated band structure shows three Dirac cones, while the pristine Bi$_2$Se$_3$
has two degenerate Dirac cones. The top-surface Dirac cone [labeled as III in Fig.2(c)] is separated
from the bottom-surface Dirac cone [labeled as I in Fig.2(b)] due to broken IS.
The top-surface (bottom-surface) cone has the binding energy of 0.812 eV (0.234 eV).
Here the binding energy is a difference between the Dirac point and $E_F$. Right below the bottom-surface
Dirac point, $E_{DP}^b$, a new Dirac cone [labeled as II in Fig.2(b)] appears with the binding energy of
0.331~eV. The similar new Dirac cone was shown in the case of K adsorption \cite{Park13}.

The binding energies of the Dirac cones can be understood from a combination of surface relaxation and charge
transfer. Let us first discuss the charge transfer effect. The amount of charge transfer is quantified from
$C_{\textmd{net}}(z)$$=C[\textmd{Bi$_2$Se$_3$/EuS}]- C[\textmd{Bi$_2$Se$_3$}] -C[\textmd{EuS}]$,
where $C[A]$ is charge of the slab geometry $A$, averaged over the $xy$-plane.
To estimate it, we perform self consistent calculations using the same supercell size and DFT parameters
and the relaxed geometry of the TI/FMI slab.
As shown in Fig.~\ref{fig:chg}(b), the values of
$C_{\textmd{net}}(z)$ for the interfacial Eu and Se atoms, Eu(1) and Se(1), are negative and positive, respectively,
suggesting that there must be a charge transfer from the EuS to the TI slab.
By integrating $C_{\textmd{net}}(z)$ from $z=0$ up to the midpoint of the Se(1) and Eu(1) positions,
~we obtain 0.092~$e$ per unit cell area, among
which 0.0042~$e$ is transferred to the bottommost QL [Fig.~\ref{fig:chg}(c)]. By filling the charge up in a Dirac
cone with the Fermi velocity $v_F=5 \times 10^5$m/s \cite{HZhang-Bi2Se3,SC_Zhang_model} from the Dirac point,
we estimate that the binding
energies of the top-surface and bottom-surface Dirac cones increase to about 0.9 eV and 0.2 eV, respectively. Then the
effect of surface relaxation gives a small decrease in the binding energy of the top-surface Dirac cone, while a small
increase in the binding energy of the bottom-surface Dirac cone.

\begin{figure}
\begin{center}
\includegraphics[width=7cm, angle=0]{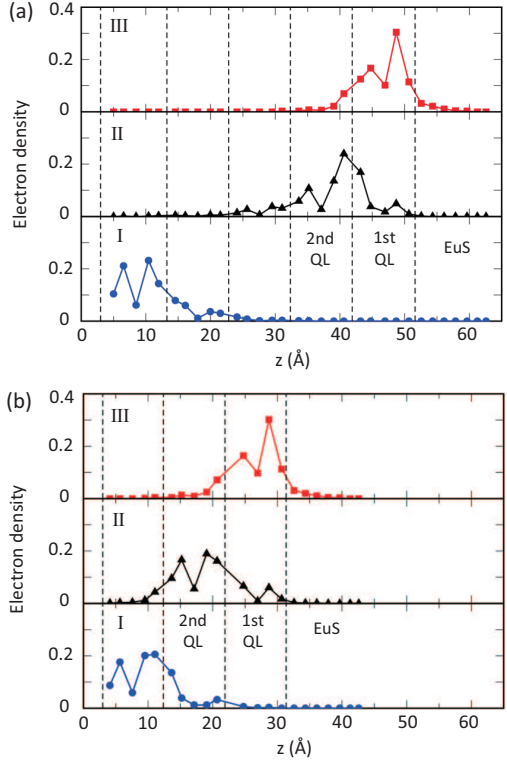}
\caption{ (Color online) (a) Calculated electron density of the top-surface Dirac cone (top panel), new Dirac
point (middle panel), and bottom-surface Dirac cone (bottom panel) at $\Gamma$ for the 5QL/EuS slab.
(b) Calculated electron density of the top-surface Dirac cone, lowest-energy quantum-well states in the bulk
conduction band region, and bottom-surface Dirac cone at $\Gamma$ for the 3QL/EuS slab.
The labels I, II, and III follow those in Fig.2. }
\label{fig:density}
\end{center}
\end{figure}

We now discuss the energy gaps in the top-surface and bottom-surface Dirac cones followed by that in the new Dirac cone.
The top-surface Dirac cone (labeled III in Fig. ~\ref{fig:band}) has an energy gap of 9.0 meV, which can be understood from the following results:
(i) a small amount of magnetic moment is induced only within the topmost QL; (ii) the top-surface states are
mostly localized into the topmost QL with only 7\% into the EuS film, as shown in Fig.~\ref{fig:density}(a). The
magnetic moments of the interfacial Eu and the atoms in the topmost QL (labeled in Fig.~\ref{fig:geo}) are as follows:
$\mu$[Eu(1)]$=$6.793, $\mu$[Se(1)]$=-$0.024, $\mu$[Bi(1)]$=$0.010, $\mu$[Se(2)]$=-$0.004, $\mu$[Bi(2)]$=$0.004,
$\mu$[Se(3)]$=$0.004~$\mu_B$. The largest induced magnetic moment comes from Se(1), and the next largest moment from Bi(1).
The fact that the largest induced moment is only $-$0.024~$\mu_B$, is consistent with a highly localized character
of Eu $f$ electrons. The magnitude of the induced moment drops rapidly as the distance from the interface
increases, and the direction of the moment alternates with the distance. The magnetic moment of the Eu(1) is slightly
smaller than that of the bulk EuS, 6.962 $\mu_B$.

Our result of the top-surface cone differs from that at the Bi$_2$Se$_3$/MnSe interface \cite{Luo-MnSe,Chulkov-1}
in three aspects. Firstly, for the latter, the states giving the largest energy gap of about 50 meV, are localized mostly
into the MnSe region and the interfacial Se atom, rather than into the topmost QL. In our case, not only the top-surface
Dirac cone is localized into the topmost QL, but also the orbitals representing the Dirac cone are similar to those for
the top-surface and bottom-surface Dirac cones in a pristine Bi$_2$Se$_3$ slab. This feature is applied to not just near
the $\Gamma$ point but in the whole $k$ space. Therefore, we claim that the top-surface
Dirac cone still persists with a small energy gap upon EuS adsorption, although it appears with a large binding energy
below the bulk valence band region. Secondly, for the Bi$_2$Se$_3$/MnSe interface, the induced magnetic moment
was the largest for the topmost Bi atom from the interface (0.04~$\mu_B$), while the magnetic moment of the topmost
Se atom was less than 0.01~$\mu_B$. Thirdly, the Bi$_2$Se$_3$/MnSe system has some ambiguity in determination of
an interface in comparison to our system, because the interfacial Se atom could be considered as part of both
Bi$_2$Se$_3$ and MnSe.

The bottom-surface Dirac cone (labeled I) has an energy gap less than 1 meV, and it is mostly localized into the bottommost QL,
as shown in Fig.~\ref{fig:density}(a). The small energy gap is due to the short-ranged induced magnetic moment from
the EuS. Interestingly, just below the bottom-surface Dirac point, an additional
Dirac cone [labeled II in Fig.2(b)] with an energy gap of 2.0 meV appears.
This new Dirac cone is localized slightly deeper into the TI slab and further away
from the EuS film than the top-surface Dirac cone, as shown in the middle panel of Fig.~\ref{fig:density}(a).
As a consequence, the magnetic moment induced into the new Dirac cone is fairly small, giving a smaller energy gap than
the top-surface Dirac cone.

\begin{table}
\begin{center}
\caption{The Fermi level and the energies (in eV) of the top-surface and bottom-surface Dirac points $E_{DP}^{t,b}$,
the new Dirac point $E_{DP}^n$, and two lowest-energy quantum-well states $E_{QWS1,QWS2}$ at $\overline{\Gamma}$ for the
$N$-QLs ($N$=3,5) with the EuS film. Here $E_{DP}^t$ is obtained from the midpoint of the gapped top-surface
Dirac cone, and $V_1=E_{DP}^b - E_{DP}^t$. Here the vacuum energy is set to zero.}
\label{table1}
\begin{ruledtabular}
\begin{tabular}{c|c|c|c|c|c|c|c}
   $N$ & $E_F$   & $E_{DP}^t$ & $E_{DP}^b$ & $V_1$ & $E_{DP}^n$ & $E_{QWS1}$ & $E_{QWS2}$ \\
   \hline
   3      & -5.222 & -6.052   & -5.639   & 0.413    &  N/A     & -5.548 & -5.250 \\
   5      & -5.258 & -6.070   & -5.493   & 0.577    & -5.589   & -5.331 & -5.204 \\
\end{tabular}
\end{ruledtabular}
\end{center}
\end{table}

We discuss features of quantum-well states (QWS) shown in the band structure Fig.~\ref{fig:band}(a). Near
the $\Gamma$ point, there are three QWS with strong Rashba spin splitting in the bulk conduction band region
and three QWS in the bulk valence band region. The QWS right above the top-surface Dirac point has an energy
gap of 2 meV at $\Gamma$, while the other five QWS do not have an energy gap. Whether the QWS are
gapped or not, is determined by their electron density profiles. As momentum $k$ increases from $\Gamma$, the
QWS in the bulk conduction band region become coupled to the bottom-surface Dirac cone, while the QWS in the
bulk valence band region are coupled to the top-surface Dirac cone. The absolute energies of two QWS in the
bulk conduction band region, QWS1 and QWS2, are listed in Table ~\ref{table1}. Interestingly, along
$\overline{\Gamma K}$, $E_F$ crosses six bands, while along $\overline{\Gamma M}$, $E_F$ crosses four
bands. The six bands are two top-surface states, a bottom-surface state, a new Dirac surface state, and
the QWS1 pair, while the four bands are the bottom-surface state, the new Dirac surface state, and the QWS1 pair.

\subsection{Effect of Bi$_2$Se$_3$ thickness}

Now we examine an effect of TI slab thickness by computing the electronic structure of the 3QL/EuS slab.
As shown in Fig.~\ref{fig:band}(d), the top-surface Dirac cone has the binding energy of 0.830~eV, which
is similar to that for the 5QL/EuS case. This result is consistent with the observation that the charge
transfer mostly occurs within the 2-3 QLs from the interface.
On the other hand, the bottom-surface Dirac cone
has the binding energy of 0.417~eV for the 3QL/EuS slab, which is about 0.2 eV larger than that for the
5QL/EuS case. For the 3QL case, the surface relaxation alone makes the bottom-surface Dirac cone more
electron-doped so that the binding energy increases up to about 0.3~eV. Then the binding energy
increases slightly more due to a small charge transfer to the bottom-most QL. The effect of surface
relaxation is more prominent for a thinner slab.

Interestingly, in contrast to the 5QL/EuS case, there are no new Dirac states for the 3QL/EuS.
This is due to the concerted effect of the charge transfer and surface relaxation. The states located
at 0.09~eV above $E_{DP}^b$ are QWS referred to as QWS1 with large Rashba spin splitting,
labeled as II in Fig. 2(e) and listed in Table ~\ref{table1}.
At small nonzero $k$ values, the QWS1 are
coupled to the bottom-surface states and open up a large gap of 78.4~meV. As $k$ increases further,
they are coupled to the top-surface states. Along $\overline{\Gamma K}$, $E_F$ crosses two top-surface
states, a bottom-surface state, one QWS1, and the QWS2 pair.

The top-surface (bottom-surface) Dirac cone has an energy gap of 9 meV (less than 1 meV), similarly to the
5-QL case, although the surface-surface hybridization is stronger for the 3-QL slab
($\Delta$ in Table II). Our result is justified
from the effective model discussed in Sec.IV.A. The QWS1 has an energy gap of 1 meV, which is slightly
smaller than that for the new Dirac cone for the 5-QL case. This difference can be due to the small difference
in the electron density profile. Compare the middle panels in Fig.~\ref{fig:density}(a) and (b). The The QWS1 for the 3-QL case is
mostly localized into the 2nd QL from the interface rather than the midpoint between the 1st and 2nd QLs.

\section{Model Hamiltonian and spin-orbital texture}

We construct a low-energy effective 4$\times$4 Hamiltonian that can explain the DFT-calculated gaps
of the surface-state Dirac cones, adapted from Ref.~\cite{Yu-QAH}. Then using this Hamiltonian, we
investigate the spin-orbital texture of the surface states, and compare it with the DFT result.

\subsection{Model Hamiltonian}

The surface states at the Bi$_2$Se$_3$/EuS interface can be described by the following effective Hamiltonian
\begin{equation}
H=
 \begin{pmatrix}
    H_R - V_t\boldsymbol{I} - \boldsymbol{M}\cdot\boldsymbol{\sigma}  &  \Delta\boldsymbol{I} \\
    \Delta\boldsymbol{I}  & - H_R - V_b\boldsymbol{I} \\
  \end{pmatrix}
  \ \ .
\label{eq:hamil}
\end{equation}

We use the basis set,
$\{|t \uparrow \rangle, |t \downarrow \rangle, |b \uparrow \rangle, |b \downarrow \rangle \}$, where
$t$ and $b$ represent top-surface and bottom-surface states, and $\uparrow$ and $\downarrow$
refer to as the
electron spin directions along the $+z$ and $-z$ axes. Here the Rashba SOC Hamiltonian
$H_R = \hbar v_{F}({\sigma}_{x}k_{y}-{\sigma}_{y}k_{x})$, where $v_F$ is the Fermi velocity and
$\boldsymbol{\sigma}=({\sigma}_{x},{\sigma}_{y},{\sigma}_{z})$
are the Pauli matrices. $\boldsymbol{I}$ is a 2$\times$2 identity matrix. The binding energy of the
top-surface (bottom-surface) Dirac cone is denoted as $V_t$ ($V_b$), where $V_b < V_t$, in order to
incorporate the charge transfer. $\boldsymbol{M}$ is the effective exchange field from the EuS.
Note that $\boldsymbol{M}$ is applied only to the top-surface states because this is a short-ranged
proximity-induced field, not an ordinary magnetic field. The coupling between the top-surface
and bottom-surface states gives an energy gap of 2$\Delta$ at $\Gamma$ for a pristine thin TI slab.

The effective Hamiltonian differs from that in Refs.~\cite{Zhang-spin}
in the sense that IS is broken via $V_t$, $V_b$, and $\boldsymbol{M}$. With an out-of-plane magnetization,
$\boldsymbol{M}=m \boldsymbol{\hat{z}}$ $(m>0)$, we find that eigenvalues of the Hamiltonian at
$\mathbf{k}=0$ are

\begin{equation}
\begin{split}
\lambda_{1,2}= \frac{1}{2}\left[ - V_2 \mp m - \sqrt{(V_1 \pm m)^2 + 4{\Delta}^{2}} \right],
\\
\lambda_{3,4}= \frac{1}{2}\left[ - V_2 \mp m + \sqrt{(V_1 \pm m)^2 + 4{\Delta}^{2}} \right],
\end{split}
\label{eq:ham}
\end{equation}
where $V_1=E_{DP}^b - E_{DP}^t$ and $V_2 \equiv V_t + V_b$.  Here $\lambda_{1,2}$ correspond
to pure top-surface states, while $\lambda_{3,4}$ pure bottom-surface states, when $\Delta=0$.

Let us now assume that $\Delta, m \ll V_1$ for the 5QL/EuS and 3QL/EuS interfaces, which is
consistent with the DFT result listed in Tables~\ref{table1} and \ref{table2}.
With this assumption, one can show that the surface state energy gaps are, up to the order
of $1/V_1^2$, written as
\begin{equation}
2m \left(1 - \frac{m^2}{4 V_1^2} - \frac{{\Delta}^{2}}{V_1^2} \right) \ ,
\ \ 2m \left( \frac{m^2}{4 V_1^2} + \frac{{\Delta}^{2}}{V_1^2}  \right) \ .
\end{equation}
Therefore, as long as $m^2/V_1^2 \ll 1$ and ${\Delta}^2 /{V_1}^2 \ll 1$ are
satisfied, the Hamiltonian dictates that the top-surface energy gap is close to 2$m$, while
the bottom-surface energy gap is negligible. In addition, the surface-state energy gaps
do not depend on TI slab thickness for slabs $\geq$3 QLs, which agrees with
the DFT calculations.

\begin{table}
\begin{center}
\caption{The hybridization gap $2 \Delta$ for the pristine $N$ QLs ($N$=3, 5), and the energy gap
of the top-surface Dirac cone for the N-QL/EuS slabs. Here $V_t=E_F-E_{DP}^t$ and $V_b=E_F-E_{DP}^b$.}
\label{table2}
\begin{ruledtabular}
\begin{tabular}{c|c|c|c|c}
$N$ & $2\Delta$ & $2m$   & $V_t$  & $V_b$   \\ \hline
   3      & 0.036    & 0.009 & 0.830 & 0.417  \\
   5      & 0.006    & 0.009 & 0.812 & 0.234  \\
\end{tabular}
\end{ruledtabular}
\end{center}
\end{table}

\subsection{Spin-Orbital Texture}

To examine the spin-orbital texture of the surface states, we consider the 5QL/EuS since $\Delta$ for the 5-QL
slab is negligible. Ignoring the surface-surface hybridization, we can describe low-energy properties of
the top (bottom) surface states using the upper (lower) 2$\times$2 block diagonal matrix in Eq. ~(\ref{eq:hamil}).
The spin-orbital texture of the bottom-surface states in our case is the same as that of a pristine TI slab
\cite{Zhang-spin} because $V_b \boldsymbol{I}$ in Eq.~(\ref{eq:hamil}) does not affect the texture. Therefore,
we first calculate the spin-orbital texture of the top-surface states using both the block diagonal matrix and
DFT. Then we present the texture of the new Dirac surface states by using DFT only. Note that the new Dirac
surface states are localized deeper into the TI slab. The DFT calculations inherently
include surface-bulk coupling, while the model Hamiltonian deals with surface states only. In this regard,
the effective Hamiltonian has a limitation to study the new Dirac surface states.

\subsubsection{Change of orbital and spin basis}

\begin{figure}
\begin{center}
\includegraphics[width=7cm, angle=0]{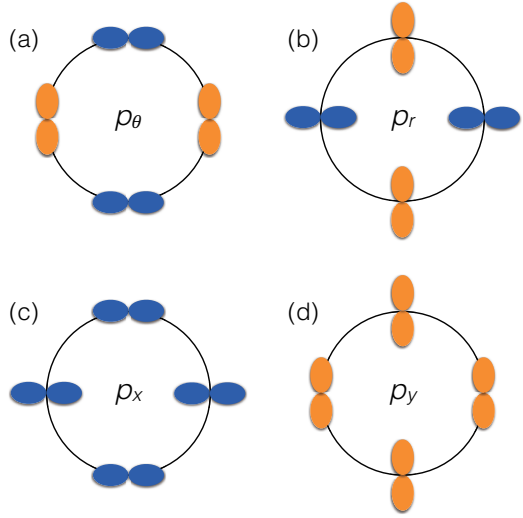}
\caption{ (Color online) Schematic diagrams showing the relationships between two different in-plane orbital basis sets:
($p_r$,$p_{\theta}$) and ($p_x$,$p_y$), where $p_r$ and $p_{\theta}$ are radial and tangential $p$ orbitals.
}
\label{fig:orb_basis}
\end{center}
\end{figure}

The upper block-diagonal matrix, $H_R - V_t\boldsymbol{I} - \boldsymbol{M}\cdot\boldsymbol{\sigma}$,
has two eigenvalues $\lambda_{\pm}=\pm \sqrt{m^2 + (\hbar v_F k)^2}$. The eigenvectors corresponding
to the upper and lower Dirac cone of the top-surface states can be, respectively, written as
\begin{eqnarray}
|\Phi_{\pm} \rangle &=&\frac{1}{N_{\pm}} \left( \pm i e^{-i\theta_k} \eta_{\pm} |\Psi_1 \rangle + |\Psi_2 \rangle \right),
\end{eqnarray}
where $\theta_k$ is the polar angle of an in-plane momentum vector ${\vec k}$,
$\eta_{\pm} \equiv \hbar v_F k/(\pm \lambda_{\pm} \pm m)$, $k \equiv \sqrt{k_x^2+k_y^2}$, and
$N_{\pm} \equiv \sqrt{1 + \eta_{\pm}^2}$. With time-reversal symmetry, $\eta_{\pm}=1$.
Considering only $p$ orbitals to the zeroth-order in the
{\bf k}$\cdot${\bf p} method, we use the basis functions such as
\begin{eqnarray}
|\Psi_1 \rangle &=&\sum_{\alpha} \left( u_{\alpha} |\alpha, p_z, \uparrow \rangle +
v_{\alpha} |\alpha, p_+, \downarrow \rangle  \right),\\
|\Psi_2 \rangle &=&\sum_{\alpha} \left( u_{\alpha} |\alpha, p_z, \downarrow \rangle
+ v_{\alpha} |\alpha, p_-, \uparrow \rangle \right).
\end{eqnarray}
Here $|p_{\pm} \rangle=\mp (|p_x \rangle \pm i |p_y \rangle)/\sqrt{2}$.
The basis functions $|\Psi_{1,2} \rangle$ form a time-reversal symmetry pair with the $z$ component
of the total angular momentum $\pm 1/2$, respectively.
Here $u_{\alpha}$ and $v_{\alpha}$ are material-dependent parameters for a pristine Bi$_2$Se$_3$ slab,
and $\alpha$ represents atom indices. The values of $u_{\alpha}$ and $v_{\alpha}$ can be obtained from
DFT calculations. However, as far as qualitative features of the spin-orbital texture are
concerned, they are not needed. Our model differs from Ref.~\cite{Zhang-spin}, in that $\eta_{\pm} \neq 1$
due to the broken time-reversal symmetry, and that the basis functions contain only the zeroth-order terms
in the {\bf k}$\cdot${\bf p} method.

Now rewriting the eigenvectors $|\Phi_{\pm} \rangle$ in terms of $p_z$ and radial and tangential $p$
orbitals ($p_r$, $p_{\theta}$), similarly to Ref.~\cite{Zhang-spin}, we find
\begin{eqnarray}
|\Phi_{+ \rangle} &=& \frac{1}{N_+} \sum_{\alpha}  (  u_{\alpha}
    \{ (\eta_+ + 1) | \alpha, p_z, \uparrow_{\theta} \rangle  +
    ( 1 - \eta_+) |\alpha,p_z, \downarrow_{\theta} \rangle \} \nonumber \\
   & & -\frac{i}{\sqrt{2}} v_{\alpha} \{ (\eta_+ + 1) |\alpha, p_r, \uparrow_{\theta} \rangle
   + (\eta_+ - 1) |\alpha, p_r, \downarrow_{\theta} \rangle \}  \nonumber \\
   & & + \frac{1}{\sqrt{2}} v_{\alpha} \{ (\eta_+ - 1) |\alpha, p_{\theta}, \uparrow_{\theta} \rangle
            + (\eta_+ + 1) |\alpha, p_{\theta}, \downarrow_{\theta} \rangle \} ) ,
\label{eq:phi1}
\end{eqnarray}
\begin{eqnarray}
|\Phi_- \rangle &=& \frac{1}{N_-} \sum_{\alpha}  ( u_{\alpha}
    \{ (\eta_- + 1) | \alpha, p_z, \downarrow_{\theta} \rangle +
    (1 - \eta_-) |\alpha, p_z, \uparrow_{\theta} \rangle \}  \nonumber \\
    & & + \frac{i}{\sqrt{2}} v_{\alpha} \{ (\eta_- + 1) |\alpha, p_r, \downarrow_{\theta} \rangle
    + (\eta_- - 1) |\alpha, p_r, \uparrow_{\theta} \rangle \}   \nonumber \\
    & & - \frac{1}{\sqrt{2}} v_{\alpha} \{ (\eta_- - 1) |\alpha, p_{\theta}, \downarrow_{\theta} \rangle
            + (\eta_- + 1) |\alpha, p_{\theta}, \uparrow_{\theta} \rangle \} ) ,
\label{eq:phi2}
\end{eqnarray}
where $|p_r \rangle = \cos\theta_k |p_x \rangle + \sin\theta_k |p_y \rangle$ and
$|p_{\theta}\rangle= -\sin\theta_k |p_x \rangle + \cos\theta_k |p_y \rangle$. Figure~\ref{fig:orb_basis}
schematically shows how the radial and tangential $p$ orbitals are related to the $p_x$ and $p_y$
orbitals. For example, the $p_x$ orbital becomes tangential at $\theta_k=\pm \pi/2$, while
it becomes radial at $\theta_k=0$ and $\pi$.
Here $| \uparrow_{\theta} \rangle=(1/\sqrt{2}) (+i e^{-i \theta_k} |\uparrow \rangle + |\downarrow \rangle)$
represents the left-handed spin texture, where the spin magnetic moment rotates clockwise
as $\theta_k$ increases.
$| \downarrow_{\theta} \rangle=(1/\sqrt{2}) (-i e^{-i \theta_k} |\uparrow \rangle + |\downarrow \rangle)$
represents the right-handed spin texture, where the spin moment rotates counter-clockwise as $\theta_k$
increases. To compare with the DFT results, we project the wave functions Eqs.~(\ref{eq:phi1}) and (\ref{eq:phi2})
onto the orbital basis, $p_x$, $p_y$, and $p_z$, and calculate the expectation values of the $x$, $y$, and $z$
components of the spin magnetic moment with respect to the projected wave functions,
$\langle \sigma_{x,y,z} \rangle_{p_x,p_y,p_z}$.

\subsubsection{Top-surface Dirac Cone}

\begin{figure}
\begin{center}
\includegraphics[width=10cm, angle=0]{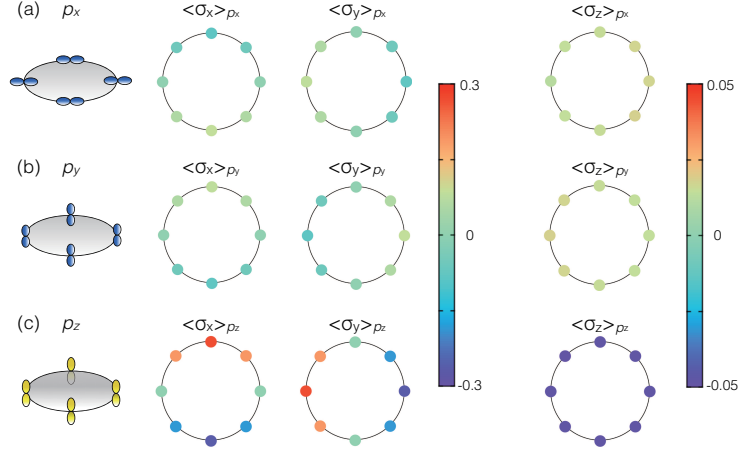}
\caption{ (Color online) (a)-(c) Spin-orbital texture of the upper Dirac cone of the top-surface states
in the Bi$_2$Se$_3$/EuS at $k=0.0094$~\AA$^{-1}$, where the leftmost column shows the schematic figures of
$p_x$, $p_y$, and $p_z$ orbitals. Note that $\langle \sigma_{x,y} \rangle$ have different scales from
$\langle \sigma_z \rangle$.}
\label{fig:spin_upper}
\end{center}
\end{figure}

We first examine the {\it upper} Dirac cone of the {\it top} surface states. The expectation values projected
onto $p_{x,y,z}$ orbitals are written as
\begin{eqnarray}
\langle \sigma_z \rangle_{p_x,p_y}&=& \sum_{\alpha} v_{\alpha}^2 \frac{1-\eta_+^2}{N_+^2}, \: \: \:
\langle \sigma_z \rangle_{p_z} = \sum_{\alpha} u_{\alpha}^2 \frac{\eta_+^2 - 1}{N_+^2},
\label{eq:sigma_1}
\end{eqnarray}
\begin{eqnarray}
\langle \sigma_x \rangle_{p_x,p_y}&=& \mp \sum_{\alpha} v_{\alpha}^2 \frac{\eta_+}{N_+^2} \sin\theta_k, \: \: \:
\langle \sigma_x \rangle_{p_z} =  \sum_{\alpha} u_{\alpha}^2 \frac{2\eta_+}{N_+^2} \sin\theta_k,
\label{eq:sigma_2}
\end{eqnarray}
\begin{eqnarray}
\langle \sigma_y \rangle_{p_x,p_y}&=& \mp \sum_{\alpha} v_{\alpha}^2 \frac{\eta_+}{N_+^2} \cos\theta_k, \: \: \:
\langle \sigma_y \rangle_{p_z} = - \sum_{\alpha} u_{\alpha}^2 \frac{2 \eta_+}{N_+^2} \cos\theta_k,
\label{eq:sigma_3}
\end{eqnarray}
where the minus (plus) sign on the left-hand side of Eqs.~(\ref{eq:sigma_2})-(\ref{eq:sigma_3}) is for
$p_x$ ($p_y$). DFT calculations are carried out at a small momentum $k$ such as 0.0094~\AA$^{-1}$, in order
to avoid hexagonal warping effect. At this $k$ point, $\eta_+=0.8654$, when we use $v_F=5 \times 10^5$~m/s
\cite{HZhang-Bi2Se3,SC_Zhang_model}. Figure~\ref{fig:spin_upper} shows the DFT-calculated spin-orbital texture.

The projected $z$ component of spin moment, Eq.~(\ref{eq:sigma_1}), does not depend on $\theta_k$,
while the in-plane components, Eqs.~(\ref{eq:sigma_2})-(\ref{eq:sigma_3}), depend on $\theta_k$.
Figure~\ref{fig:spin_upper} shows that $\langle \sigma_x \rangle_{p_x,p_y,p_z}$ is antisymmetric about
the $x$ axis, while $\langle \sigma_y \rangle_{p_x,p_y,p_z}$ is antisymmetric about the $y$ axis.
Thus, the projected $x$ and $y$ components of spin moment exactly follow the $\sin\theta_k$ and
$\cos\theta_k$ dependence, respectively. Because $0 < \eta_+ < 1$, interestingly,
$\langle \sigma_z \rangle_{p_z} < 0 < \langle \sigma_z \rangle_{p_x,p_y}$,
as suggested from Eq.~(\ref{eq:sigma_1}) and shown in the last column of Fig.~\ref{fig:spin_upper}.

Let us discuss the helicity of the in-plane spin texture coupled to $p_x$, $p_y$, and $p_z$ orbitals.
For $p_z$ orbital, $\langle \sigma_{x} \rangle_{p_z}$ in Fig.~\ref{fig:spin_upper}(c) shows that the
$x$ component of spin moment points along the $+x$ axis at $\theta_k=\pi/2$, and along the $-x$ axis
at $\theta_k=3\pi/2$. $\langle \sigma_{y} \rangle_{p_z}$ in Fig.~\ref{fig:spin_upper}(c) shows that the
$y$ component of spin moment points along the $-y$ axis at $\theta_k=0$, and the $+y$ axis at
$\theta_k=\pi$. Thus, the spin moment coupled to the $p_z$ orbital rotates clockwise,
which implies left-handed spin-texture. The spin texture coupled to $p_x$ and $p_y$ orbitals
can be, similarly, understood from Figs. ~\ref{fig:spin_upper}(a) and (b) combined with Fig. ~\ref{fig:orb_basis}.
Using $\langle \sigma_{x} \rangle_{p_x}$ and $\langle \sigma_{y} \rangle_{p_y}$
in Fig.~\ref{fig:spin_upper}, we find that the tangential $p$ orbital has right-handed spin
texture. Similarly, we find that the radial $p$ orbital has left-handed spin texture.

Combining the result discussed so far, we discuss the overall spin-orbital texture.
The DFT result gives
$|\langle \sigma_z \rangle_{p_x,p_y}|/|\langle \sigma_z \rangle_{p_z}|
=\sum_{\alpha} v_{\alpha}^2 / \sum_{\alpha} u_{\alpha}^2 = 0.3094$. As a result, the $p_z$ orbital
contributes dominantly to $\langle \sigma_z \rangle$ over the $p_x$ and $p_y$ orbitals.
When this ratio is applied to Eqs.~(\ref{eq:sigma_2})-(\ref{eq:sigma_3}), we find that the contributions
of the $p_z$ orbital to $\langle \sigma_{x,y} \rangle$ are dominant over those of the $p_x$ and $p_y$ orbitals.
In our case, $\eta_+$ is close to unity because $m \ll |\hbar v_F k|$ near $\Gamma$. Thus,
Eqs.~(\ref{eq:sigma_1})-(\ref{eq:sigma_3}) imply that $|\langle \sigma_z \rangle_{p_z}|$ is smaller than
the maximum value of $|\langle \sigma_{x,y} \rangle_{p_z}|$. This agrees with the DFT result,
Fig.~\ref{fig:spin_upper}(c). Overall, the surface states in the upper Dirac cone have a strong left-handed
in-plane spin texture with a small out-of-plane spin moment along the negative $z$ axis, as shown in Fig.~\ref{fig:spin_upper}(c).

Our result differs from the spin-resolved ARPES data on Mn-doped Bi$_2$Se$_3$ (Figs. 3(e)-(g) in Ref.~\cite{XU12}),
where the spin texture was observed in the presence of a strong external magnetic field. Firstly, the
Bi$_2$Se$_3$ slab in the Bi$_2$Se$_3$/EuS interface responds diamagnetically to the weak exchange field from the EuS,
while the Mn-doped Bi$_2$Se$_3$ does not have diamagnetic response to an external magnetic field.
Note that the magnetization of the EuS film aligns along the positive $z$ axis. The out-of-plane spin moment
along the negative $z$ axis is due to diamagnetic response of Bi$_2$Se$_3$ to the ferromagnetic EuS film.
The diamagnetic nature of Bi$_2$Se$_3$ has been shown in experiments \cite{Gupta,HLi}.
Secondly, and spin-orbital texture was not examined in the previous experiment \cite{XU12}.

\begin{figure}
\begin{center}
\includegraphics[width=10cm, angle=0]{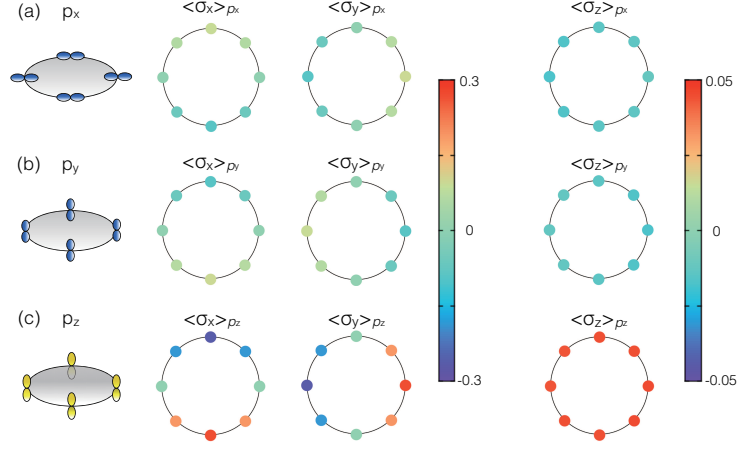}
\caption{(Color online) (a)-(c) Spin-orbital texture of the lower Dirac cone of the top-surface states
in the Bi$_2$Se$_3$/EuS at $k=0.0094$~\AA$^{-1}$.}
\label{fig:spin_lower}
\end{center}
\end{figure}

Next, we examine the spin-orbital texture of the {\it lower} Dirac cone of the {\it top} surface states.
In this case, the texture is the opposite to that of the upper Dirac cone discussed earlier.
The expectation values are given as
\begin{eqnarray}
\langle \sigma_z \rangle_{p_x,p_y}&=&\sum_{\alpha} v_{\alpha}^2 \frac{1-\eta_-^2}{N_-^2}, \: \: \:
\langle \sigma_z \rangle_{p_z} = \sum_{\alpha} u_{\alpha}^2 \frac{\eta_-^2 - 1}{N_-^2},
\label{eq:sigma_4}
\end{eqnarray}
\begin{eqnarray}
\langle \sigma_x \rangle_{p_x,p_y}&=& \pm \sum_{\alpha} v_{\alpha}^2 \frac{\eta_-}{N_-^2} \sin\theta_k, \: \: \:
\langle \sigma_x \rangle_{p_z} = - \sum_{\alpha} u_{\alpha}^2 \frac{2\eta_-}{N_-^2} \sin\theta_k,
\label{eq:sigma_5}
\end{eqnarray}
\begin{eqnarray}
\langle \sigma_y \rangle_{p_x,p_y}&=& \pm \sum_{\alpha} v_{\alpha}^2 \frac{\eta_-}{N_-^2} \cos\theta_k, \: \: \:
\langle \sigma_y \rangle_{p_z} =   \sum_{\alpha} u_{\alpha}^2 \frac{2 \eta_-}{N_-^2} \cos\theta_k,
\label{eq:sigma_6}
\end{eqnarray}
where the plus (minus) sign on the left-hand side of Eqs.~(\ref{eq:sigma_5})-(\ref{eq:sigma_6}) is for $p_x$ ($p_y$).
DFT calculations are performed at $k=0.0094$~\AA$^{-1}$, and at this $k$ point, $\eta_-=1.1556$ using
$v_F=5 \times 10^5$~m/s \cite{HZhang-Bi2Se3,SC_Zhang_model}. Figure~\ref{fig:spin_lower} shows the DFT-calculated spin-orbital texture.

Because of $\eta_- > 1$, $\langle \sigma_z \rangle_{p_x,p_y} < 0$ and $\langle \sigma_z \rangle_{p_z} > 0$ from
Eq.~(\ref{eq:sigma_4}) and shown in the last column of Fig.~\ref{fig:spin_lower}. This feature is the opposite to
that in Fig.~\ref{fig:spin_upper}. The signs of $\langle \sigma_x \rangle_{p_x,p_y,p_z}$
($\langle \sigma_y \rangle_{p_x,p_y,p_z}$) in Fig.~\ref{fig:spin_lower} or Eqs.~(\ref{eq:sigma_5})-(\ref{eq:sigma_6})
are reversed to those in Fig.~\ref{fig:spin_upper} or Eqs.~(\ref{eq:sigma_2})-(\ref{eq:sigma_3}) about the $x$ axis
($y$ axis). Therefore, the $p_z$ and radial (tangential) orbitals are now coupled to the right- (left-)handed
spin texture. Since the contributions of the $p_z$ orbital are dominant over the $p_x$ and $p_y$ orbitals, overall,
the surface states have right-handed in-plane spin texture with the out-of-plane spin moment along the
positive $z$ axis.

\subsubsection{Bottom-surface Dirac Cone}

The proximity-induced effect does not influence the bottom-surface Dirac cone.
Thus, $\eta_{\pm}=1$,
and Eqs.~(\ref{eq:sigma_1})-(\ref{eq:sigma_6}) give $\langle \sigma_z \rangle_{p_x,p_y,p_z}=0$. Note that
the bottom-surface Dirac cone is gapless within numerical accuracy. For the upper (lower) Dirac cone of
the bottom-surface states, the $p_x$, $p_y$, and $p_z$ orbitals are coupled to the in-plane spin texture
in the same fashion as those for the lower (upper) Dirac cone of the top-surface states. This agrees
with our DFT calculations (not shown).

\subsubsection{New Dirac Cone}

We also examine the the spin-orbital texture of the new Dirac cone appearing in the 5 QLs with EuS.
The upper (lower) Dirac cone of the new
surface states shows qualitatively similar spin-orbital texture to the upper (lower) Dirac cone of the top
surface states. Three small quantitative differences are as follows. Compared to the top-surface states,
(i) $|\langle \sigma_{x,y} \rangle_{p_z}|$ increases by 0.048-0.059$\mu_B$;
(ii) $\langle \sigma_{x,y} \rangle_{p_x,p_y}$ decreases by 0.019-0.028$\mu_B$;
(iii) $\langle \sigma_{z} \rangle_{p_z}$ and $\langle \sigma_{z} \rangle_{p_x,p_y}$  decrease
by 0.022-0.028 and 0.007-0.012$\mu_B$, respectively,
where the two different numbers for each difference come from the upper and lower Dirac
cone. The differences in the $x$ and $y$ components of the spin moment arise because the orbitals of the new
surface states slightly differs from those of the top-surface states. The new surface states have larger
contributions from the $p_z$ orbital and smaller contributions from the $p_x$ and $p_y$ orbitals
than the top-surface states. The difference in the $z$ component originates from the fact that the new Dirac
surface states are localized slightly deeper into the TI slab, relative to the top and bottom-surface states,
as shown in Fig.~\ref{fig:density}. As a consequence, the proximity effect of the EuS film is weaker on the
new Dirac surface states than on the top-surface states.

\section{Summary}

In summary, we investigated the magnetic proximity effect on the electronic structure and spin-orbital texture
of the Dirac surface states from the Bi$_2$Se$_3$/EuS slab through first-principles calculations and the effective model.
The Dirac surface states localized into the QL right next to the interface, open up an energy gap of 9~meV, independently of the TI slab
thickness for slabs as thick as 3 QLs or beyond. However, the Dirac surface states localized into the other side
of the interface, remains gapless. These features of the gaps are due to the short-ranged induced magnetic moments
into the TI slab. For the 5QL/EuS slab, we found that a new Dirac cone was formed with an energy
gap of 2 meV, while there was no such new Dirac cone for the 3QL/EuS slab. We constructed the effective model Hamiltonian
which includes surface-surface interaction, magnetic proximity effect, and band bending, in order to explain the gap of
the top and bottom-surface Dirac cones. By setting the spin-orbital basis for the model Hamiltonian, we computed
the spin-orbital texture with broken time reversal symmetry and this calculated result agree with the DFT calculations.

\section{Acknowledgments}
This work was supported by the National Institute of Supercomputing and Networking and
Korea Institue of Science and Technology Information with supercomputing resources including
technical support (KSC-2013-C2-023), as well as Virginia Tech Advanced Research Computing.
The authors are grateful to Bart Partoens from University of Antwerp for stimulating discussion
on similar magnetic topological insulators. K.P. was supported by U.S. National Science
Foundation DMR-1206354.

\end{document}